\documentclass[12pt]{cernart}
\usepackage[dvips]{graphics}
\usepackage{multicol}

\begin{document}

\begin{titlepage}
\docnum{CERN--EP/02--??}
\date{19.12.2002}

\title{\Large \bf \boldmath Search for the decay $\mathbf{\mathit{K_S}}\rightarrow\pi^0\gamma\gamma$}

\begin{Authlist}
{\bf NA48 Collaboration} \\[0.4cm]

A.~Lai, D.~Marras \\
{\small \em Dipartimento di Fisica dell'Universit\`a e Sezione
    dell'INFN di Cagliari, I-09100 Cagliari, Italy.} \\[0.4cm]

R.~Batley, A.~Bevan, R.S.~Dosanjh, 
T.J.~Gershon, G.E.~Kalmus\footnote{Based at Rutherford Appleton 
Laboratory, Chilton, Didcot,OX11 0QX, U.K.},
D.J.~Munday, E.~Olaiya, 
M.A.~Parker, T.~White, S.A.~Wotton \\
{\small \em Cavendish Laboratory, University of Cambridge, 
    Cambridge, CB3 0HE, U.K\footnote{Funded by the U.K.
    Particle Physics and Astronomy Research Council.}.} \\[0.4cm]

G.~Barr, G.~Bocquet, A.~Ceccucci,
T.~Cuhadar-D\"{o}nszelmann, D.~Cundy, N.~Doble,
V.~Falaleev, L.~Gatignon, A.~Gonidec,
B.~Gorini, G.~Govi,
P.~Grafstr\"om, W.~Kubischta, A.~Lacourt,
M.~Lenti\footnote{On leave from Sezione dell'INFN di Firenze, 
I-50125 Firenze, Italy.}, A.~Norton, S.~Palestini, 
B.~Panzer-Steindel,
G.~Tatishvili\footnote{On leave from Joint Institute for Nuclear 
Reaserch, Dubna, 141980, Russian Federation}, H.~Taureg, 
M.~Velasco, H.~Wahl \\
{\small \em CERN, CH-1211 Geneva 23, Switzerland.} \\[0.4cm]

C.~Cheshkov, P.~Hristov\footnote{Present address: EP, 
CERN, CH-1211, Geneva 23, Switzerland}, V.~Kekelidze, 
D.~Madigojine, N. Molokanova,
Yu.~Potrebenikov, A.~Tkatchev, A.~Zinchenko \\
{\small \em Joint Institute for Nuclear Research, Dubna, Russian
    Federation.}  \\[0.4cm] 

I.~Knowles, C.~Lazzeroni, V.~Martin, 
R.~Sacco, A.~Walker \\
{\small \em Department of Physics and Astronomy, University of
    Edinburgh, JCMB King's Buildings, Mayfield Road, Edinburgh,
    EH9 3JZ, U.K.} \\[0.4cm]

M.~Contalbrigo, P.~Dalpiaz, J.~Duclos,
P.L.~Frabetti, A.~Gianoli, M.~Martini,
L.~Masetti, F.~Petrucci, M.~Savri\'e \\
{\small \em Dipartimento di Fisica dell'Universit\`a e Sezione
    dell'INFN di Ferrara, I-44100 Ferrara, Italy.} \\[0.4cm]

A.~Bizzeti\footnote{Dipartimento di Fisica dell'Universita' di Modena e Reggio Emilia, 
via G. Campi 213/A I-41100 Modena, Italy}, M.~Calvetti, G.~Collazuol,
G.~Graziani, E.~Iacopini \\
{\small \em Dipartimento di Fisica dell'Universit\`a e Sezione
    dell'INFN di Firenze, I-50125 Firenze, Italy.} \\[0.4cm]

H.G.~Becker, M.~Eppard, H.~Fox,
K.~Holtz, A.~Kalter, K.~Kleinknecht,
U.~Koch, L.~K\"opke, P.~Lopes da Silva,
P.~Marouelli, I.~Pellmann, A.~Peters,
B.~Renk, S.A.~Schmidt, V.~Sch\"onharting,
Y.~Schu\'e, R.~Wanke, A.~Winhart, M.~Wittgen \\
{\small \em Institut f\"ur Physik, Universit\"at Mainz, D-55099
    Mainz, Germany\footnote{Funded by the German Federal Minister for
    Research and Technology (BMBF) under contract 7MZ18P(4)-TP2.}.} \\[0.4cm] 

J.C~Chollet, L.~Fayard, L.~Iconomidou-Fayard,
J.~Ocariz, G.~Unal, I.~Wingerter \\
{\small \em Laboratoire de l'Acc\'el\'eratur Lin\'eaire, IN2P3-CNRS,
Universit\'e de Paris-Sud, 91898 Orsay,
France\footnote{Funded by Institut National de Physique des Particules
et de Physique Nucl\'eaire (IN2P3), France}.} \\[0.4cm]

G.~Anzivino, P.~Cenci, E.~Imbergamo,
P.~Lubrano, A.~Mestvirishvili, A.~Nappi, M.~Pepe, M.~Piccini \\
{\small \em Dipartimento di Fisica dell'Universit\`a e Sezione
    dell'INFN di Perugia, I-06100 Perugia, Italy.} \\[0.4cm]

R.~Carosi, R.~Casali, C.~Cerri, M.~Cirilli,
F.~Costantini, R.~Fantechi, S.~Giudici,
I.~Mannelli, G.~Pierazzini, M.~Sozzi \\
{\small \em Dipartimento di Fisica dell'Universit\`a, Scuola Normale 
Superiore e Sezione dell'INFN di Pisa, I-56100 Pisa, Italy.} \\[0.4cm]

J.B.~Cheze, J.~Cogan, M.~De Beer, P.~Debu,
A.~Formica, R.~Granier de Cassagnac, E.~Mazzucato,
B.~Peyaud, R.~Turlay, B.~Vallage \\
{\small \em DSM/DAPNIA - CEA Saclay, F-91191 Gif-sur-Yvette Cedex,
    France.} \\[0.4cm]

M.~Holder, A.~Maier, M.~Ziolkowski \\
{\small \em Fachbereich Physik, Universit\"at Siegen, D-57068 
Siegen, Germany\footnote{Funded by the German Federal Minister for
Research and Technology (BMBF) under contract 056SI74.}.} \\[0.4cm]

R.~Arcidiacono, C.~Biino, 
N.~Cartiglia, R.~Guida, 
F.~Marchetto, E.~Menichetti, N.~Pastrone \\
{\small \em Dipartimento di Fisica Sperimentale dell'Universit\`a e
    Sezione dell'INFN di Torino, \\ I-10125 Torino, Italy.} \\[0.4cm]

J.~Nassalski, E.~Rondio, M.~Szleper,
W.~Wislicki, S.~Wronka \\
{\small \em Soltan Institute for Nuclear Studies, Laboratory for High
    Energy Physics, \\ PL-00-681 Warsaw, Poland\footnote{
    Supported by the Committee for Scientific Research grants 5P03B10120, 2P03B11719 and SPUBM/CERN/P03/DZ210/2000 and using computing resources of the Interdisciplinary Center for Mathematical and Computational Modelling of the University of Warsaw.}.}\\[0.4cm]

H.~Dibon, G.~Fischer, M.~Jeitler,
M.~Markytan, I.~Mikulec, G.~Neuhofer,
M.~Pernicka, A.~Taurok, L.~Widhalm \\
{\small \em \"Osterreichische Akademie der Wissenschaften, Institut
    f\"ur Hochenergiephysik, \\ A-1050 Wien, Austria.} \\[0.4cm]

\end{Authlist}

\vskip 1.0truecm
\submitted{\small (Submitted to {\it Physics Letters B})}
\vskip 1.0truecm


\newpage

\begin{abstract}
A search for the decay $K_S\rightarrow\pi^0\gamma\gamma$ has been made using the NA48 detector at the CERN SPS. Using data collected in 1999 during a 40-hour run with a high-intensity $K_S$ beam, an upper limit for the branching ratio $BR\left(K_S\rightarrow\pi^0\gamma\gamma, z\ge 0.2\right)<3.3\times\nolinebreak 10^{-7}$ has been obtained at 90\% confidence level, where $z=m_{\gamma\gamma}^2/m_{K_0}^2$.
\end{abstract}

\end{titlepage}

\section{Introduction}
The study of the decays $K_{S,L}\rightarrow\pi^0\gamma\gamma$ is useful to test the predictions of the Chiral Perturbation Theory $(\chi PT)$ on branching ratios and $q^2$-spectra \cite{Ecker1}. 
At present the branching ratio for the decay $K_L\rightarrow\pi^0\gamma\gamma$ has been measured precisely \cite{klpgg}, while the corresponding $K_S$ decay has not yet been observed. 
$\chi PT$ predicts a branching ratio for $K_S\rightarrow\pi^0\gamma\gamma$ of $3.8\times 10^{-8}$ with the cut-off $z=m_{\gamma\gamma}^2/m_{K_0}^2\ge 0.2$, to avoid the region with the pion pole which dominates the total rate \cite{Ecker1}.
The amplitude can be calculated precisely as it is non-vanishing in lowest order Chiral Perturbation Theory and expected to be quite insensitive to higher order corrections.
With sufficient statistics, the chiral structure of weak vertices could be tested experimentally in the high-$q^2$ region from a comparison with the predicted $q^2$-spectrum.

\section{Experimental set-up and data taking}
The NA48 detector, originally designed to measure the direct CP-violation parameter $\epsilon'/\epsilon$ in neutral kaon decays with simultaneous $K_L$ and $K_S$ beams, was used to collect data for studies on rare $K_S$ decays during a test run in 1999 with a high-intensity $K_S$ beam. 
This beam had an intensity about 200 times higher than the one used in the $\epsilon'/\epsilon$ runs.
The present analysis is based on data collected during this 40-hour high-intensity $K_S$ run.
Neutral kaons were produced by a primary beam of $450\: \mathrm{GeV}/c$ protons, extracted from the SPS accelerator during a 2.4 s spill every 14.4 s and impinging with an angle in the vertical plane of 4.2 \nolinebreak mrad on a 2 mm diameter, 400 mm long beryllium target, at an intensity of $\sim6\times 10^9$ protons per pulse.
The target was followed by a sweeping magnet and a 0.36 \nolinebreak cm diameter collimator to define a narrow beam of neutral kaons and hyperons.
The beginning of the fiducial volume was set 500 cm downstream of the centre of the target, 1 m before the end of the collimator.
The fiducial volume was contained in an 89 m long, evacuated steel cylinder and followed by the main NA48 detector.
The neutral particles of the beam were contained in a 156 mm diameter carbon-fibre pipe linked to the evacuated cylinder and passing through the detector, in order to avoid interactions of neutrons and photons from the beam.
A detailed description of the detector and beam lines can be found elsewhere \cite{e'}.

In the present analysis the following sub-detectors were used: 
a magnetic spectrometer composed of four drift chambers, two upstream and two downstream of a dipole magnet, for vetoing charged particles;
seven anti-counter rings (AKL) of iron and plastic scintillator, for vetoing photons outside the acceptance of the main detector;
a quasi-homogeneous electromagnetic liquid-krypton calorimeter (LKr), to measure energy, position and time of electromagnetic showers generated by photons;
a sampling hadron calorimeter composed of 96 steel and scintillator planes.
The electromagnetic calorimeter is composed of 13212 readout tower cells of $2\times2\: \mathrm{cm}^2$ cross section and $27\:X_0$ depth each. 
The energy resolution is \cite{Calo}
\begin{equation}
\frac{\sigma(E)}{E}\simeq\frac{0.09}{E}\oplus\frac{0.032}{\sqrt{E}}\oplus0.0042,
\end{equation}
with E in GeV. 
The position and time resolutions for a single photon are better than 1 \nolinebreak mm and 500 ps, respectively, for energies greater than 25 GeV.
The overall energy scale was determined by a fit of the position of an anti-counter (AKS) used during $\epsilon'/\epsilon$ runs.

Events satisfying the trigger conditions for $\gamma\gamma$ or $3\pi^0$ events were selected, since no dedicated trigger for $\pi^0\gamma\gamma$ (and $2\pi^0$) decays was set up during the special high-intensity $K_S$ run.
The trigger decisions were based on quantities reconstructed from orthogonal projections of the energy deposit in the electromagnetic calorimeter \cite{Fischer}.
The trigger for both $\gamma\gamma$ and $3\pi^0$ decays required:
\begin{enumerate}
\item the total deposited energy $E_{tot}=E_{LKr}+E_{HAD}$ being larger than 50 GeV;
\item the centre of gravity of the event, computed from the first moments $M_{1,x}$ and $M_{1,y}$ of the energy in each projection,
\begin{equation}
c.o.g^{trig}=\frac{\sqrt{M_{1,x}^2+M_{1,y}^2}}{E_{LKr}},
\end{equation}
being less than 15 cm from the beam axis;
\item the decay vertex of the $K^0$, computed from the second moments $M_{2,x}$ and $M_{2,y}$, 
\begin{equation}
z_{vertex}^{trig}=z_{LKr}-\frac{\sqrt{E_{LKr}\left(M_{2,x}+M_{2,y}\right)-\left(M_{1,x}^2+M_{1,y}^2\right)}}{m_K},
\label{zverttrig}
\end{equation}
being less than 15 m downstream from the end of the collimator. This corresponds to about 2.5 $K_S$ lifetimes, since the $K_S$ have a mean energy of 110 GeV. In equation \ref{zverttrig}, $z_{LKr}$ is the $z$ coordinate of the electromagnetic calorimeter with respect to the target and $m_K$ is the kaon mass.
\end{enumerate}
The $3\pi^0$ trigger required also at least 4 energy peaks in each projection. The cuts on the vertex for the $\gamma\gamma$ trigger and on the maximum number of energy peaks for the $3\pi^0$ trigger were varied during the run, while the downscaling factors were adjusted according to the obtained rate. 
The different trigger conditions and downscaling factors are summarised in Table \ref{trig}.

\begin{table}
\begin{small}
\begin{center}
\begin{tabular}{|c|cc|cc|c|}
\hline
 & Cond. $\gamma\gamma$ & $D_{\gamma\gamma}$ & Cond. $3\pi^0$ & $D_{3\pi^0}$ & $N_{2\pi^0}$\\
\hline
a & zv $<12$ m; no peak cut & 3 & npx $>4$ or npy $>4$ or npx$=$npy$=$4 & 1 & $\sim 5\times 10^4$\\
b & zv $<9$ m; no peak cut & 3 & npx $>4$ or npy $>4$ or npx$=$npy$=$4 & 1 & $\sim 3.4\times 10 ^5$ \\
c & zv $<9$ m; no peak cut & 3 & npx $>4$ or npy $>4$ & 1 & $\sim 3.4\times 10^5$\\
d & zv $<9$ m; no peak cut & 2 & npx $>4$ or npy $>4$ & 1 & $\sim 1.5\times 10^6$\\
e & zv $<9$ m; npx,npy$=$1,2 & 1 & npx $>4$ or npy $>4$ & 1 & $\sim 1.1\times 10^6$\\
\hline
\end{tabular}
\end{center}
\end{small}
\caption{\label{trig} Conditions and downscaling factors $D$ for $\gamma\gamma$ and $3\pi^0$ triggers, for different parts of the run.  zv$=z_{vertex}^{trig}-6$ m is the position of the decay vertex starting from the beginning of the decay region, npx and npy the number of energy peaks in x and y projection respectively, in a time window of 20 ns. In the rightmost column the number of $2\pi^0$ events after all selection cuts satisfying $\gamma\gamma$ or $3\pi^0$ trigger conditions is given.}
\end{table}

The trigger efficiency for $K_S\rightarrow 2\pi^0$ decays was computed using a minimum bias trigger which did not involve the electromagnetic calorimeter: for the first part of the run (conditions a-d in Table \ref{trig}) it was ($98.52\pm0.06$)\%, while it decreased to ($22.0\pm0.2$)\% for condition e.
About one third of the statistics used for this analysis was collected under this last condition.
The trigger efficiency for $K_S\rightarrow\pi^0\gamma\gamma$ events was assumed to be the same as for $2\pi^0$ events. Possible differences were studied from simulation and considered as systematic uncertainties. 

\section{Event selection}
To select $K_S\rightarrow\pi^0\gamma\gamma$ candidates, events with at least four electromagnetic clusters were considered. Each cluster was required to satisfy the following conditions:
\begin{enumerate}
\item the energy had to be greater than 3 GeV and less than 100 GeV;
\item each cluster had to be well inside the calorimeter acceptance, more than 15 cm and less than 113 cm from the axis of the beam pipe;
\item the distance from any dead cell of the calorimeter had to be greater than 2 cm;
\item the distance from any other cluster in the event with an energy greater than 1.5 \nolinebreak GeV had to be greater than 10 cm.
\end{enumerate}
For each event, all combinations of four clusters passing the pre-selection described above were considered. For each combination, the following selection criteria were applied: 
\begin{enumerate}
\item the time of each cluster ($t_{cl}$) had to be within $\pm 3\:\mathrm{ns}$ of the event time ($t_{av4}$) computed as the average of $t_{cl}$ weighted over the energy;
\item no other cluster (not belonging to the combination considered) with an energy greater than 1.5 GeV had to be found within $\pm 5\:\mathrm{ns}$ of the event time, to reduce the effect of accidental background;
\item the sum of the four cluster energies had to be greater than 70 GeV and less than 170 GeV;
\item in order to distinguish $\pi^0\gamma\gamma$ decays from $2\pi^0$ decays, a $\chi^2$ variable was defined \cite{epsi}:
\begin{equation}
\chi_{2\pi^0}^2=\left[\frac{\left(m_{\gamma\gamma}+m'_{\gamma\gamma}\right)/2-m_{\pi^0}}{\sigma_+}\right]^2+\left[\frac{\left(m_{\gamma\gamma}-m'_{\gamma\gamma}\right)/2}{\sigma_-}\right]^2,
\end{equation}
where $m_{\gamma\gamma}$ and $m'_{\gamma\gamma}$ are the invariant masses of the two $\gamma \gamma$ pairs and $\sigma_{\pm}$ the resolutions of $\left(m_{\gamma\gamma}\pm m'_{\gamma\gamma}\right)/2$ measured from the data.
For each combination of four clusters the $\gamma\gamma$ pairs giving the lowest $\chi^2$ were considered and, if more than one combination of four clusters satisfied the first three requirements, the one with the lowest $\chi^2$ was selected. 
\end{enumerate}
Further requirements were then applied to each event:
\begin{enumerate} 
\item to define a good event, cuts in the centre of gravity and decay vertex position of the kaon were used. The centre of gravity
\begin{equation}
c.o.g.=\frac{\sqrt{\left(\sum_i{E_ix_i}\right)^2+\left(\sum_i{E_iy_i}\right)^2}}{\sum_i{E_i}},
\end{equation} 
with $E_i$, $x_i$, $y_i$ respectively the energy, x and y coordinates in the LKr for the $i$-th cluster, had to be less than 7 cm;
\item the decay vertex
\begin{equation}
z_{vertex}=z_{LKr}-\frac{1}{m_{K^0}}\sqrt{\sum_{i,j<i}{E_iE_j\cdot\left[\left(x_i-x_j\right)^2+\left(y_i-y_j\right)^2\right]}},
\end{equation}
was required to be within the first 9 m of the decay volume, in order to reject background from $K_L\rightarrow3\pi^0$ events with missing or overlapped photons;
\item events with signals in the drift chambers within $\pm10$ ns of the event time or with overflows in chamber 1, 2 or 4 within $\pm250$ ns were rejected. The overflow condition is generated in the drift chamber readout whenever more than seven hits in a plane are detected within 100 ns, in order to avoid recording events with high hit multiplicity;
\item in order to reject events with overlapping showers, an energy-dependent cluster-width cut was applied, using a parametrisation taken from $K_S\rightarrow2\pi^0$ decays;
\item events with signals in any of the AKL rings within $\pm3$ ns of the event time were rejected;
\item if the energy deposited in the hadron calorimeter was greater than 3 GeV, events were rejected if any HAC signal was found in a time window of $\pm15$ ns around the event time. 
\end{enumerate}

\section{Background rejection}
Background to $K_S\rightarrow\pi^0\gamma\gamma$ may come from misreconstructed $K_S\rightarrow2\pi^0$ with four clusters in the acceptance; from $K_S\rightarrow2\pi^0$ with three clusters in the acceptance and one random cluster accidentally in time; from $K_S\rightarrow\pi^0\pi^0_D$ with one lost particle and one misidentified charged particle; from $K_L\rightarrow3\pi^0$ with two lost or overlapped photons; and from $K_L\rightarrow\pi^0\gamma\gamma$.

\subsection{Background from misreconstructed $K_S\rightarrow2\pi^0$}
\begin{figure}
\begin{center}
\resizebox{15.5cm}{!}{\includegraphics{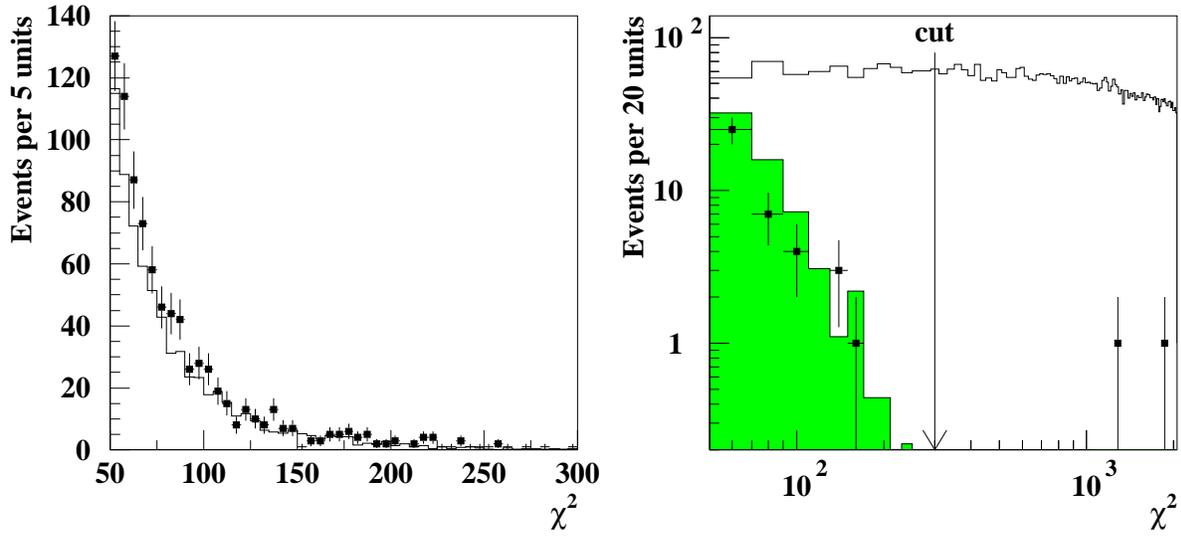}}
\end{center}
\caption{\label{contr_reg} Left: $\chi^2_{2\pi^0}$ distribution for data (squares) and background simulation (solid lines) in $\chi^2_{2\pi^0}$ control region, before the cut on \textit{z}. Distributions are normalised to the $K_S$ flux.
Right: $\chi^2_{2\pi^0}$ distribution for data (squares), background simulation for misreconstructed $K_S\rightarrow2\pi^0$ (filled histogram) and $K_S\rightarrow \pi^0\gamma\gamma$ simulation (solid line). For the $K_S\rightarrow \pi^0\gamma\gamma$ simulation the normalisation is arbitrary.}
\end{figure}

To define a good $\pi^0\gamma\gamma$ event, the $\gamma\gamma$ pair with invariant mass closest to the nominal $\pi^0$ mass was required to be within $\pm 3\:\mathrm{MeV}/c^2$ of the nominal $\pi^0$ mass, corresponding to about 3 standard deviations of the resolution.
$K_S\rightarrow 2\pi^0$ events with misreconstructed cluster energy due to non-Gaussian tails in the energy resolution were simulated and found to be concentrated in a region with $\chi^2_{2\pi^0}$ below 300. Therefore a $\chi^2_{2\pi^0}$ greater than 300 was required for the $\pi^0\gamma\gamma$ candidates.
In order to compare data and simulation a control region with $50<\chi^2_{2\pi^0}<300$ was defined and the distribution in $\chi^2_{2\pi^0}$ is shown in Fig.~\ref{contr_reg} left.
The cut on $\chi^2_{2\pi^0}$ discards about 8\% of signal events (Fig.~\ref{contr_reg} right).
The remaining $\gamma\gamma$ pair had to satisfy the requirement $z=m_{\gamma\gamma}^2/m_{K_0}^2\ge 0.2$, in order to reject background from $2\pi^0$ events and to be able to compare the result with the prediction made by $\chi PT$, which is given with the cut-off $z>0.2$.
After all cuts, 2 $K_S\rightarrow\pi^0\gamma\gamma$ candidates were selected. i

From Monte Carlo simulation the background from $K_S\rightarrow2\pi^0$ decays with four clusters in the acceptance is expected to be less than 0.5 events at 90\% C.L., which corresponds to a rejection factor of at least $4\times 10^7$. 

\subsection{Background from $K_S\rightarrow2\pi^0$ with three clusters in the acceptance and one random cluster in time}
\begin{figure}
\begin{center}
\resizebox{15.5cm}{!}{\includegraphics{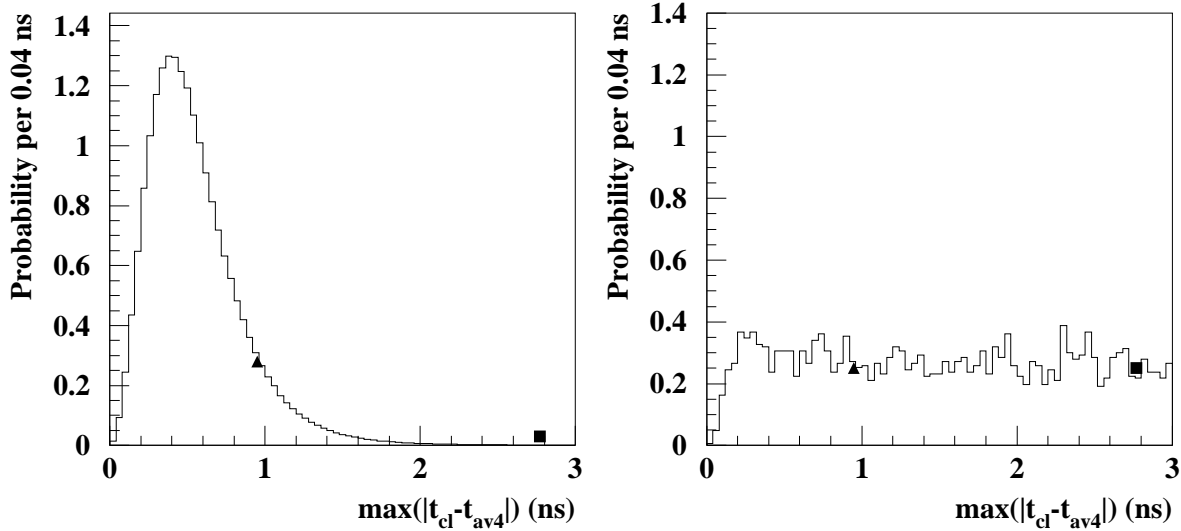}}
\end{center}
\caption{\label{like} $\left|t_{cl}-t_{av4}\right|$ for the farthest in-time cluster from the event time in $2\pi^0$ decays in the data (left) and for the random cluster in background simulation (right). For the random cluster a flat distribution in $\left|t_{cl}-t_{av3}\right|$ was assumed. The full square and triangle represent $\max\left(\left|t_{cl}-t_{av4}\right|\right)$ for each of the two selected events.} 
\end{figure}

Using simulated $K_S\rightarrow2\pi^0$ events and events in the data selected by a random trigger, the background from $K_S\rightarrow2\pi^0$ decays with three clusters in the acceptance and one random cluster accidentally in time was estimated to be $3.0\pm0.1_{stat}$ events. 
Random events were assumed to have a flat distribution in time with respect to the average time of the 3 clusters of the $2\pi^0$ event ($t_{av3}$) and were overlayed to the simulated $K_S\rightarrow2\pi^0$ with three clusters in the acceptance to reconstruct a $K_S\rightarrow\pi^0\gamma\gamma$ candidate.
One of the 2 selected $K_S\rightarrow\pi^0\gamma\gamma$ candidates in the data contains one cluster for which the time is compatible with the distribution of the random cluster in background simulation.  
For the other $K_S\rightarrow\pi^0\gamma\gamma$ candidate, the farthest in-time cluster from the event time has the same probability to be a random cluster or to be part of an event with four photons in the acceptance of the calorimeter (Fig.~\ref{like}).

\subsection{Background from $K_S\rightarrow\pi^0\pi^0_D$ and $K_L\rightarrow3\pi^0$} 
Applying all the selection criteria described above to simulated $K_S\rightarrow\pi^0\pi^0_D$ and $K_L\rightarrow3\pi^0$ events, no candidate remains, corresponding to a rejection factor of at least $2.5\times 10^6$ for $K_S\rightarrow\pi^0\pi^0_D$ and of at least $7.5\times 10^5$ for $K_L\rightarrow3\pi^0$. The background from $K_S\rightarrow\pi^0\pi^0_D$ is expected to be less than 0.2 events, while the background from $K_L\rightarrow3\pi^0$ is expected to be less than 0.1 events.

\subsection{Background from $K_L\rightarrow\pi^0\gamma\gamma$}
A non reducible source of background is given by the decay $K_L\rightarrow\pi^0\gamma\gamma$, which cannot be distinguished from the signal. Using the branching fraction measured by the NA48 experiment \cite{klpgg}, the background from this channel was computed to be about 0.1 events. 

\section{Normalisation and acceptance evaluation}
\begin{figure}
\begin{center}
\resizebox{15.5cm}{!}{\includegraphics{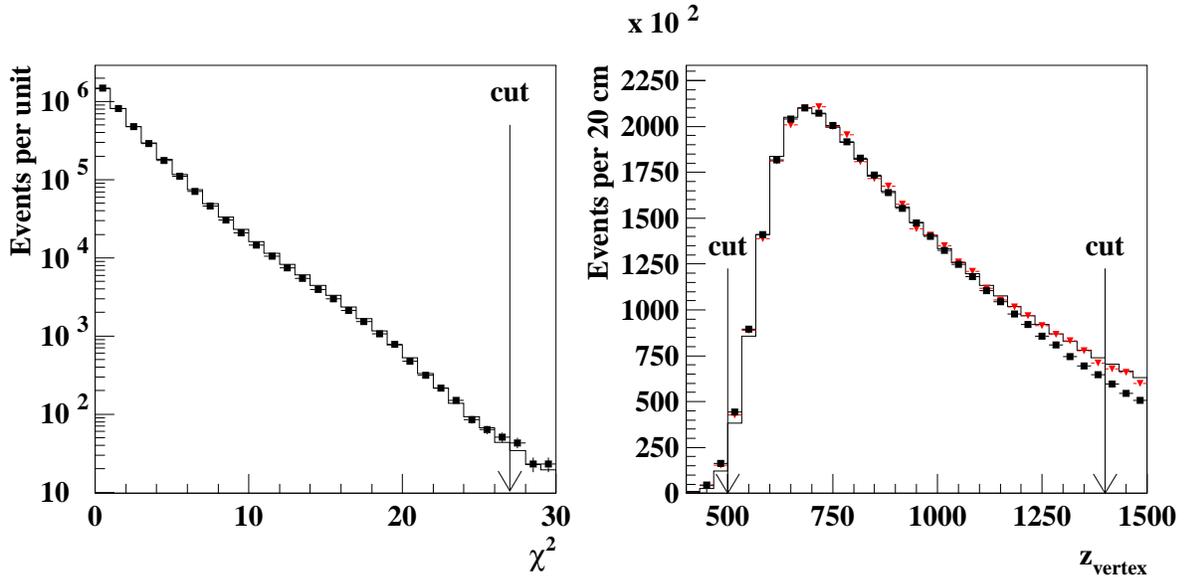}}
\end{center}
\caption{\label{MC2pi0} $\chi^2_{2\pi^0}$ (left) and $z_{vertex}$ (right) distributions of the selected $K_S\rightarrow 2\pi^0$ events for data (squares) and simulation (solid line). In the left plot the complete selection has been applied except the $\chi^2_{2\pi^0}$ cut and the distributions are normalised to the total number of events. In the right plot the complete selection has been applied except the $z_{vertex}$ cut and the distributions are normalised to the bin with the maximum number of events. The data distribution corrected for the trigger inefficiency (triangles) shows a better agreement with the simulation at high values of $z_{vertex}$.} 
\end{figure}

The decay $K_S\rightarrow2\pi^0$ was chosen as the normalisation channel. Since the two decays differ only in the invariant mass of one $\gamma\gamma$ pair and were collected by the same trigger, systematic uncertainties are expected to cancel in the ratio.
The same selection criteria were applied, with the only exceptions of $\chi^2_{2\pi^0}$, which had to be less than 27, and the invariant masses of the $\gamma\gamma$ pairs, which had to be within $\pm 3 \:\mathrm{MeV}/c^2$ of the nominal $\pi^0$ mass, considering the pairing which gives the lowest $\chi^2$. 
After all cuts, $3.3\times 10^6$ $K_S\rightarrow \nolinebreak 2\pi^0$ decays were selected, with an acceptance of 17.3\% computed from a Monte Carlo sample, which included non-Gaussian tails in the energy resolution of the calorimeter, but had no accidentals and no simulation of the trigger. 
The comparison between data and Monte Carlo simulation in the $\chi^2_{2\pi^0}$ and $z_{vertex}$ distributions for selected $K_S\rightarrow 2\pi^0$ events is shown in Fig.~\ref{MC2pi0}. At high values of $z_{vertex}$ a slight inefficiency appears in the data due to the decay vertex cut at trigger level. Correcting for this inefficiency, data and simulation are in better agreement at high $z_{vertex}$. The correction is, however, not applied to the data in the analysis, since the trigger inefficiency is assumed to be the same for signal and normalisation channel.

The total number of $K_S$ with an energy between 70 and 170 GeV and decaying in the first 8 m of the decay region was measured from $K_S\rightarrow2\pi^0$ events to be $(4.1\pm0.2)\times10^8$, taking into account the losses of statistics due to trigger downscaling ($\sim 27\%$) and to rejection of events with accidental in-time hits in the drift chambers or in the AKL ($\sim 55\%$).
This flux was also computed, as a cross check, using the decay $K_L\rightarrow3\pi^0$, which however is statistically less significant, and found to be $(3.9\pm0.4)\times10^8$. The two results are compatible within the error, which is due to the statistics used to compute acceptance and trigger efficiency and to the uncertainty on the branching ratios.
The total number of $K_S$ is not used in the limit computation, since the same trigger efficiency is assumed for signal and normalisation channel.      

\begin{figure}
\begin{center}
\resizebox{!}{7.0cm}{\includegraphics{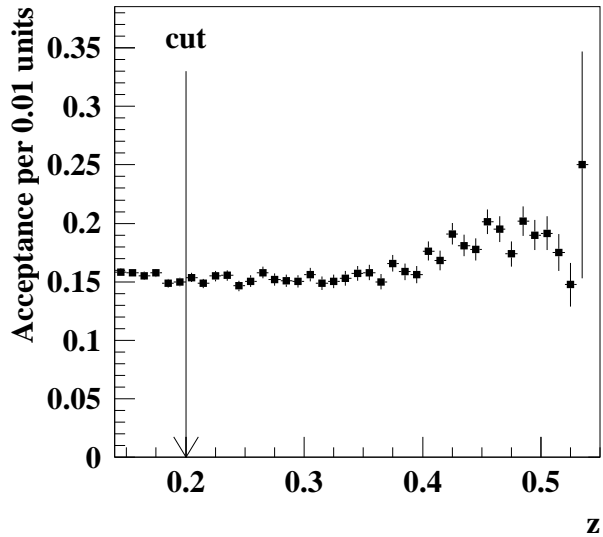}}
\end{center}
\caption{\label{acceptance} Acceptance as a function of $z=m_{\gamma\gamma}^2/m_{K^0}^2$ for $K_S\rightarrow\pi^0\gamma\gamma$ simulated events.}
\end{figure}

For the signal an acceptance of 15.6\% was determined from simulation where the amplitude predicted in $\chi PT$ \cite{Ecker1} was used.
The acceptance as a function of \textit{z} is shown in Fig.~\ref{acceptance}. 
The cut on $z\ge 0.2$ is expected to introduce a negligible systematic uncertainty, since the acceptance for the signal is nearly flat in that region of z.

\section{Systematic uncertainties}
The following effects were studied:
\begin{enumerate}
\item the background estimation was checked with an independent method based only on data. 
The background from $K_S\rightarrow2\pi^0$ decays with three clusters in the acceptance and one random cluster accidentally in time was assumed to have a flat distribution in the region 3 ns $<\left|t_{cl}-t_{av4}\right|<$ 6 ns for the farthest in-time cluster and to be the only class of background to be found in that region after applying most of the selection criteria. 
The expected background was then extrapolated from the sidebands to the signal region and compared with the expectation given by the simulation, applying the same selection.
The two results are compatible within the statistical error and the relative difference of 24\% is computed as systematic uncertainty.
\item the different cut on $\chi^2_{2\pi^0}$ between signal and normalisation, together with a slightly different resolution of the best reconstructed $\pi^0$ mass between data and simulation, produced a systematic uncertainty of 0.6\% on the ratio of acceptances when a window of $\pm 3\:\mathrm{MeV}/c^2$ is used for the cut on the $\gamma\gamma$ invariant masses;
\item from simulation the relative difference in the trigger efficiency between $K_S\rightarrow\pi^0\gamma\gamma$ and $K_S\rightarrow2\pi^0$ was computed to be about 1.7\%;
\item not including the effect of non-Gaussian tails in the simulation produced an increase in the single acceptances of about 5\%, while the variation in the ratio was 1.5\%;
\item the uncertainty of 0.9\% on the $K_S\rightarrow2\pi^0$ branching ratio gives a systematic uncertainty of the same amount;
\item the statistical error on the simulated sample is about 0.5\%;
\item varying the selection cuts had a negligible effect on the result. 
\end{enumerate}

The total systematic uncertainty was computed as the sum in quadrature of the individual uncertainties and estimated to be 24.1\%. The effect on the result is an increase of the BR limit of about 6\% \cite{Cousins}.

\section{Result and conclusions}
Applying all the selection criteria, 2 $K_S\rightarrow\pi^0\gamma\gamma$ candidates were found and \linebreak $3.1\pm\nolinebreak 0.1_{stat}\pm\nolinebreak 0.7_{syst}$ background events were expected, mainly from $K_S\rightarrow2\pi^0$ decays with three clusters in the acceptance and one random cluster accidentally in time. The acceptance was computed to be $\alpha_{\pi^0\gamma\gamma}=15.6\%$ for the signal and $\alpha_{2\pi^0}=17.3\%$ for the normalisation channel, in which $N_{2\pi^0}=3.3\times10^6$ events were selected. The same trigger efficiency was used for the two decays. Using $N_{\pi^0\gamma\gamma}<3.14$, which corresponds to a confidence level of 90\% when 2 events are seen and $3.1$ background events are expected with a systematic uncertainty of 24\% \cite{Cousins,Feld-Cous}, an upper limit for the branching ratio of the decay $K_S\rightarrow\pi^0\gamma\gamma$ was calculated:
\begin{displaymath}
BR\left(K_S\rightarrow\pi^0\gamma\gamma,z\ge 0.2\right)=\frac{N_{\pi^0\gamma\gamma}}{N_{2\pi^0}}\cdot\frac{\alpha_{2\pi^0}}{\alpha_{\pi^0\gamma\gamma}}\cdot BR\left(K_S\rightarrow2\pi^0\right)<3.3\times10^{-7}\;\; \mathrm{at \: 90\% \, C.L.}
\end{displaymath}
with $z=\frac{m_{\gamma\gamma}^2}{m_{K_0}^2}$.

This is the first upper limit on this branching ratio to be published. It lies about one order of magnitude higher than the branching ratio predicted by Chiral Perturbation Theory.

\section*{Acknowledgements}
It is a pleasure to thank the technical staff of the participating laboratories, universities and affiliated computing centres for their efforts in the construction of the NA48 apparatus, in the operation of the experiment, and in the processing of the data.

\newpage

\end{document}